# Vertical alignment of stagnation points in pseudo-plane ideal flows

Che Sun[1,2]

[1] *Institute of Oceanology, Chinese Academy of Sciences, 7 Nanhai Road, Qingdao, 266071, China*

[2] *National Laboratory for Marine Science and Technology, 1 Wenhai Road, Qingdao, 266037, China*

(Corresponding email: csun@qdio.ac.cn)

**Key words:** Ideal fluid; Pseudo-plane flow; Stagnation point; Vertical alignment; Topological fluid dynamics, Morse theory.

# Content

**Abstract.** Recent studies of pseudo-plane ideal flow (PIF) reveal a ubiquitous presence of vortex alignment in both homogeneous and stratified fluids, and in both inertial and rotating reference frames as well. The exact solutions of a steady-state PIF model suggest that stagnation points tend to be vertically aligned and the concentric structure represents a fixed-point phenomenon of the Euler equations. Exception occurs in the rotating frame when a flow holds inertial period and skew center becomes possible. Properties of stagnation points based on Morse theory are obtained, leading to a topological explanation of vertical alignment via pressure Hessian. The study thus uncovers a new aspect of vortex behavior in ideal fluid that requires vortex center to align with the direction of gravity when vortex evolution reaches a laminar end state characterized by steady pseudo-plane velocities. Though the phenomenon arises from the constraint of the Euler equations, under specific conditions the topological theory is applicable to viscous fluid and explains the curvilinear tilting of von Kármán swirling vortex.

## I. Introduction

Diagnoses of atmospheric and oceanic flows have revealed a basic state called Gravest Empirical Mode (GEM) in which a scalar property, such as temperature, has invariant vertical profile along a streamfunction contour (*Sun and Watts* 2001, *Sun* 2005). The temperature data from a baroclinic current fluctuate around the basic GEM profile with small rms residues. The low-dimensional structure motivated *Sun* (2008) to develop a steady pseudo-plane ideal flow (PIF) model based on the notion that geostrophic flows are quasi-horizontal and stratified turbulence tends to collapse into a laminar end state with negligible vertical velocity. Such pseudo-plane flows have vertically varying horizontal velocities but no vertical velocity (*Saccomandi* 1994).

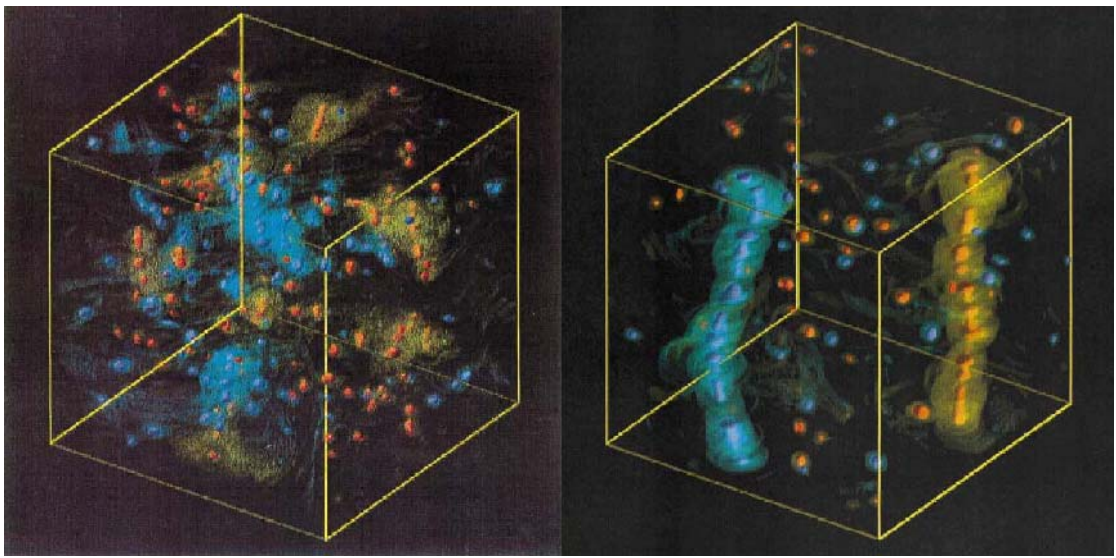

Figure 1. Vortex configurations at the initial turbulence stage (left pane) and the laminar end state (right panel) of a numerical simulation of geostrophic turbulence decay by McWilliams et al. (1999). Colors indicate positive and negative values of potential vorticity. Reproduced with permission from J. Fluid Mech. 401, 1-26 (1999). Copyright 1999 Cambridge University Press.

Meanwhile numerical experiments show that geostrophic vortices in a freely evolving turbulence have an intriguing tendency toward vertical alignment (*Polvani* 1991, *Nof and Dewar* 1994, *Viera* 1995, *McWilliams et al.* 1999). The phenomenon, as depicted in Figure 1, is different from the equivalent-barotropic structure of the GEM field and represents a weak form of vertical coherence with wider presence in geophysical fluids. Compared with the alignment of vorticity and strain rate in small-scale turbulence (*Ashurst et al.* 1987), the macroscale vortex alignment is less well-known and a theoretical explanation of its physical mechanism remains elusive.

Previous experiments reveal that evolution to an aligned state involves transient processes such as filamentation and radiation of gravity waves, but do not explain why the system shall evolve to that state. The question has to be addressed by a theoretical analysis of the end state. Recent studies of exact solutions to the PIF model shed new light on this aspect: except for inertial cases, the steady PIF solutions display universal concentric formulation in both rotating and non-rotating fluids. That is, the horizontal position of stagnation point is fixed and vertical alignment of stagnation points occurs in elliptic, hyperbolic and multipolar strain flows (*Sun* 2016, 2017). The concentric structure has obvious connection with the vortex-alignment phenomenon, because stagnation point is the common feature of steady vortices and pseudo-plane velocity characterizes the laminar end state of vortex evolution.

We intend to develop a topological theory for the concentric structure of the PIF solutions, and thereby provide a theoretical explanation for the vortex-alignment phenomenon. As it turns out, the phenomenon stems from the Euler equations and has omnipresence beyond geophysical flows.

## II. Steady pseudo-plane ideal flow

*Sun* (2008) derived the PIF model from the stratified-turbulence equations of *Riley et al.* (1981) by taking the limit of vanishing vertical velocity under low Froude number. For a steady pseudo-plane flow $\mathbf{u} = [u(x,y,z), v(x,y,z), 0]$, the Euler equations under Boussinesq approximation also reduce to the PIF model

$$uu_x + vu_y - fv = -p_x \quad (1)$$

$$uv_x + vv_y + fu = -p_y \quad (2)$$

$$\rho = -p_z \quad (3)$$

$$u_x + v_y = 0 \quad (4)$$

$$u\rho_x + v\rho_y = 0 \quad (5)$$

Here $p$ is the pressure perturbation divided by a mean density $\rho_0$, $\rho$ is the density perturbation scaled by $\rho_0/g$, and $f$ is the Coriolis parameter in a reference frame rotating with constant angular velocity ($f = 0$ for the inertial reference frame). The continuity equation (4) for incompressible fluid leads to a streamfunction $\psi(x,y,z)$ satisfying $u=-\psi_y,\ v=\psi_x$. While $\rho \neq 0$ signifies a baroclinic flow, $\rho=0$ describes homogeneous fluid in which density equation (5) is redundant. The overdetermination of the PIF model and its symmetry implication have been examined by *Sun* (2008).

In differential topology, Morse theory enables us to analyze the topology of a manifold by studying differentiable functions on that manifold (*Morse* 1925). The Hessian matrix as a square matrix of second-order partial derivatives plays an important role in Morse theory. For pseudo-plane flows, the Hessian matrix deals with function $\psi(x,y)$ of two variables and its determinant, called the Hessian, is written as $H\psi = \psi_{xx}\psi_{yy} - \psi_{xy}^2$.

The following concepts of flow topology will be used.

**Equivalent barotropic**. A pseudo-plane flow is equivalent-barotropic (EB) if its horizontal velocity vector does not change direction vertically and helicity density vanishes everywhere. Helicity measures the knottedness of vortex lines and also measures velocity rotation in a pseudo-plane flow (*Sun* 2008).

**Critical point**. A point at which all the first partial derivatives of a function are zero. The critical point is nondegenerate (degenerate) if the Hessian is non-zero (zero), and is **isolated** if it has a neighborhood containing no other critical points.

**Stagnation point**. A point with zero flow velocity. In pseudo-plane flows, stagnation point is a critical point of streamfunction. It is a relative extremum when $H\psi$ is positive, and a saddle when $H\psi$ is negative.

The Morse lemma states that nondegenerate critical points are isolated. The lemma is a sufficient condition but not a necessary one, because a degenerate critical point may also be isolated. A nondegenerate critical point in non-divergent flows is either a center or a saddle, depending on the sign of its Hessian. But a degenerate critical point can be any topologic type, including cusp (*Brøns and Hartnack* 1999).

In contrast to previous studies of flow topology that examined node, focus and saddle (*Tobak and Peake* 1982, *Perry and Chong* 1987, *Aref and Brøns* 1998, *Heil et al.* 2017), this study addresses vertical alignment of stagnation points in pseudo-plane flows and is not concerned with their exact topological type, as long as they are isolated. Examination of vertical alignment is only meaningful for isolated critical points.

## III. Examples of exact solutions

We first provide some examples of exact solutions to the PIF model and examine their stagnation points. As listed in Table I, four solutions are rotating fluid and the other five are non-rotating fluid. The derivation of these solutions can be found in *Sun* (2016, 2017).

**Table I.** List of exact PIF solutions used for the stagnation-point analysis.

| | | |
|---|---|---|
| rotating frame | S1 | Inertial circular vortex |
| | S2 | Inertial elliptic / hyperbolic flow |
| | S3 | Baroclinic elliptic / hyperbolic flow |
| | S4 | Baroclinic straightline jet |
| inertial frame | S5 | Quadratic circular vortex |
| | S6 | Elliptic / hyperbolic flow |
| | S7 | Quadratic straightline jet |
| | S8 | Quartic circular vortex |
| | S9 | Multipolar strain flow |

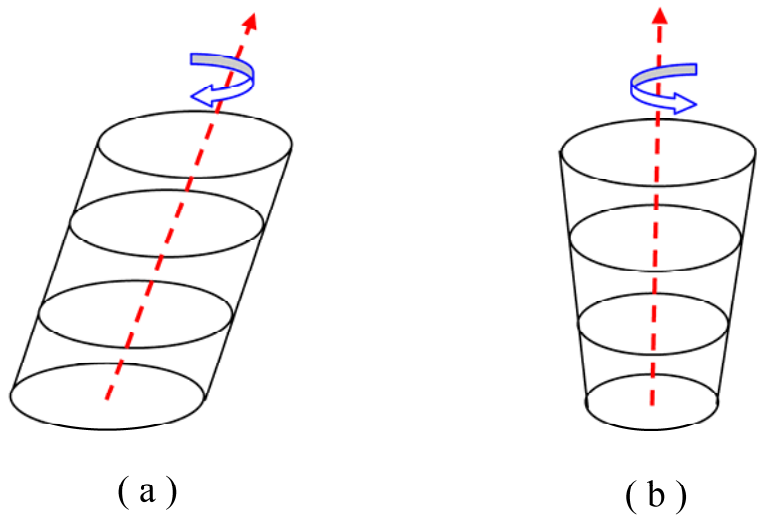

(a)  (b)

**Figure 2.** Streamline illustration for (a) circular vortex S1 with skew center and (b) circular vortices S5 and S8 with aligned center.

**S1**. Inertial circular vortex

$$\psi = -\frac{f}{2}[(x-z)^2 + y^2]$$
$$u = fy, \quad v = f(z-x)$$
$$p = 0, \quad \rho = 0$$
$$H\psi = f^2, \quad Hp = 0$$

The flow does not incur pressure perturbation. At each depth, $(z, 0)$ is an isolated stagnation point with positive Hessian, and vertically forms a tilting line as illustrated in Figure 2.

**S2.** Inertial elliptic / hyperbolic flow

$$\psi = -a(x-z)^2 - \frac{f}{2}y^2$$
$$u = fy, \quad v = -2a(x-z)$$
$$p = f(a - \frac{f}{2})y^2, \quad \rho = 0$$
$$H\psi = 2af, \quad Hp = 0$$

The solution describes an inertial elliptic ($af > 0$) or hyperbolic ($af < 0$) flow with skew center $(z, 0)$, and the *x*-axis is a critical line of pressure.

**S3.** Baroclinic elliptic / hyperbolic flow

$$\psi = a(z)x^2 + b(z)y^2$$
$$u = -2by, \quad v = 2ax$$
$$p = 2ab(x^2 + y^2) + f\psi, \quad \rho = -\frac{f(a-b)'}{a-b}\psi$$
$$H\psi = 4ab, \quad Hp = 4ab(2a+f)(2b+f)$$

where for elliptic flow

$$b = \frac{c - f \operatorname{arctanh}(h)}{2h}, \quad a = bh^2$$

and for hyperbolic flow

$$b = \frac{c - f \arctan(h)}{2h}, \quad a = -bh^2$$

For the elliptic case, $H\psi = 4b^2h^2 > 0$, the origin is a relative extremum; For the hyperbolic case, $H\psi = -4b^2h^2 < 0$, the origin is a saddle. The solution shows it is possible to have pressure saddle point in rotating fluid: for $f > -2a > 0$ in the hyperbolic case and $-2b > f > -2a > 0$ in the elliptic case, we find $Hp < 0$, which indicates a pressure saddle point at the origin.

**S4**. Baroclinic straightline jet

$$\psi = (y - z)^2$$
$$u = 2(z - y), \quad v = 0$$
$$p = f\psi, \quad \rho = 2f(y - z)$$
$$H\psi = 0, \quad Hp = 0$$

The streamfunction and pressure of this EB flow have the same critical line $y = z$ that shifts horizontally with depth. Because $\rho_y = 2f$, no density critical point exists.

**S5.** Quadratic circular vortex

$$\psi = \frac{1}{2} z(x^2 + y^2)$$
$$u = -zy, \quad v = zx$$
$$p = z\psi, \quad \rho = -2\psi$$
$$H\psi = z^2, \quad Hp = z^4$$

The positive Hessian values suggest the origin is an isolated critical point for both streamfunction and pressure.

**S6**. Elliptic and hyperbolic flows

$$\psi = a(z)x^2 + \frac{c}{a(z)}y^2$$

$$u = -\frac{2c}{a}y, \quad v = 2ax$$

$$p = 2c(x^2 + y^2), \quad \rho = 0$$

$$H\psi = 4c, \quad Hp = 16c^2$$

The flow is elliptic for $c > 0$ and hyperbolic for $c < 0$. With non-zero Hessian, the origin is an isolated stagnation point and a pressure center.

**S7**. Quadratic straightline jet

$$\psi = (x - zy)^2$$

$$u = 2z(x - zy), \quad v = 2(x - zy)$$

$$p = 0, \quad \rho = 0, \quad H\psi = Hp = 0$$

The jet direction rotates vertically. The non-isolated stagnation points at each depth form a critical line $x = zy$.

**S8**. Quartic circular vortex

$$\psi = \frac{1}{2}z(x^2 + y^2)^2$$

$$u = -2zy(x^2 + y^2), \quad v = 2zx(x^2 + y^2)$$

$$p = \frac{2}{3}z^2(x^2 + y^2)^3, \quad \rho = -\frac{4}{3}z(x^2 + y^2)^3$$

$$H\psi = 12z^2(x^2 + y^2)^2, \quad Hp = 80z^4(x^2 + y^2)^4$$

The origin is the only stagnation point at each depth, and is an isolated degenerate critical point for both streamfunction and pressure.

**S9**. Multipolar strain flow

$$\psi = r^n \sin(n\theta + Z), \quad Z = Z(z), \quad n > 2$$
$$v_r = -n r^{n-1} \cos(n\theta + Z), \quad v_\theta = n r^{n-1} \sin(n\theta + Z)$$
$$p = -\frac{1}{2} n^2 r^{2(n-1)}, \quad \rho = 0$$
$$H\psi = -n^2 (n-1)^2 r^{2(n-2)}, \quad Hp = n^4 (n-1)^2 (2n-3) r^{4(n-2)}$$

The solution was obtained by *Sun* (2016) in cylindrical coordinates and the Hessian is calculated from a polar-coordinate form given by *Arnott and Marston* (1989). At each depth the origin is the only critical point for streamfunction and pressure, and therefore is degenerate isolated.

Table II classifies the critical points of all solution examples. Those isolated stagnation points are further divided into four categories in Table III according to their density fields and pressure critical points.

**Table II**. Classification of exact solutions based on types of critical point for streamfunction, pressure and density. Non-isolated critical points form critical line and critical plane, and non-critical means there is no critical point.

|  | $\psi$ | $p$ | $\rho$ |
|---|---|---|---|
| nondegenerate isolated | S1, S2, S3 S5, S6 | S3, S5, S6 | S3, S5 |
| degenerate isolated | S8, S9 | S8, S9 | S8 |
| critical line | S4, S7 | S2, S4 |  |
| critical plane |  | S1, S7 | S1, S2, S6 S7, S9 |
| non-critical |  |  | S4 |

**Table III**. Flow types at an isolated stagnation point based on density field (homogeneous/baroclinic) and pressure critical point (nondegenerate/degenerate).

|  | nondegenerate $Hp \neq 0$ | degenerate $Hp = 0$ |
|---|---|---|
| homogeneous $\rho = 0$ | Type I ( S6 ) | Type II ( S1, S2, S9 ) |
| baroclinic $\rho \neq 0$ | Type III ( S3, S5 ) | Type IV ( S8 ) |

## IV. Properties of stagnation point

We restrict our attention to those pseudo-plane flows with isolated stagnation points. Because viscosity is absent in the PIF model, pseudo-plane flows at different depths can only interact via pressure. The key step of geometric analysis, therefore, is to find the critical-point relation between streamfunction and pressure. There are three lemmas on this regard.

**Lemma 1**. *Stagnation points in a steady pseudo-plane flow must be critical points of pressure.*

By definition there is no motion at a stagnation point. Substitution of $u = v = 0$ into the horizontal momentum equations (1-2) yields $p_x = p_y = 0$, which means the point is a pressure critical point. Unsteady flow does not have this property due to the presence of non-zero tendency terms.

**Lemma 2**. *An isolated stagnation point is an isolated critical point for any property that functionally depends on streamfunction.*

We use phenomenological approach to prove the lemma. In pseudo-plane flows there are three types of isolated critical point: center, saddle and cusp. While no isoline passes through a center, at least two isolines with different tangent directions meet at a saddle or a cusp. Conversely, if two or more isolines intersect at a point with different tangent directions, the point is a saddle or a cusp, and belongs to isolated critical point.

Assuming a property is a function of $\psi$, a streamline would correspond to an isoline of the property. Then the property has the same isoline pattern as $\psi$ at an isolated stagnation point, and belongs to one of the three types of isolated critical points. This proves the lemma.

In a PIF flow, density is either constant or a function of $\psi$. In both cases an isolated stagnation point corresponds to a density critical point as $\rho_x = \rho_y = 0$. It should be emphasized that Lemma 2 only applies to isolated stagnation points: non-isolated stagnation points, such as S4, may not correspond to density critical points.

**Lemma 3**. *A degenerate stagnation point must correspond to degenerate pressure critical point. A stagnation point corresponding to nondegenerate pressure critical point must be isolated. In the inertial reference frame, a nondegenerate stagnation point must correspond to a*

*pressure extremum.*

Taking the second-order partial differentiation of Eqs.(1-2) and using the no-flow condition at the stagnation point, we obtain

$$p_{xx} = -u_x^2 - v_x u_y + f v_x = \psi_{xx}\psi_{yy} - \psi_{xy}^2 + f\psi_{xx} = H\psi + f\psi_{xx}$$
$$p_{yy} = -u_y v_x - v_y^2 - f u_y = \psi_{xx}\psi_{yy} - \psi_{xy}^2 + f\psi_{yy} = H\psi + f\psi_{yy}$$
$$p_{xy} = -u_y u_x - v_y u_y + f v_y = f\psi_{xy}$$

in which Eq.(4) has been used to simplify $p_{xy}$. Now we can calculate the pressure Hessian at the stagnation point as

$$Hp = p_{xx}p_{yy} - p_{xy}^2 = (H\psi + f\psi_{xx})(H\psi + f\psi_{yy}) - f^2\psi_{xy}^2$$
$$= (H\psi)^2 + f(\psi_{xx} + \psi_{yy})H\psi + f^2 H\psi$$
$$= H\psi (H\psi + f\zeta + f^2)$$

which gives

$$Hp = (H\psi)^2 + f\zeta_a H\psi \qquad (6)$$

Here $\zeta_a = \zeta + f$ is absolute vertical vorticity. If the stagnation point is degenerate, i.e., $H\psi = 0$, equality (6) gives $Hp = 0$. Conversely, if the pressure critical point is nondegenerate, i.e., $Hp \neq 0$, equality (6) requires $H\psi \neq 0$, which means the stagnation point is isolated.

For non-rotating fluid or rotating fluid with $\zeta_a = 0$, equality (6) is simplified to

$$Hp = (H\psi)^2 \qquad (7)$$

which means a pressure critical point has the same Hessian degeneracy as the stagnation point. If $H\psi \neq 0$, equality (7) requires $Hp > 0$, meaning nondegenerate stagnation point in the inertial frame must correspond to a pressure extremum. The above lemma is thus proved.

Table IV. Relation between streamfunction Hessian and pressure Hessian at a stagnation point.

|  | rotating frame $f \neq 0$ | inertial frame $f = 0$ |
|---|---|---|
| nondegenerate $H\psi \neq 0$ | $Hp = 0$ ( S1, S2 ) $Hp \neq 0$ ( S3 ) | $Hp > 0$ ( S5, S6 ) |
| degenerate $H\psi = 0$ | $Hp = 0$ ( S4 ) | $Hp = 0$ ( S7, S8, S9 ) |

Together the three lemmas describe critical-point relation between different variables. The Hessian relation between streamfunction and pressure, as summarized in Table IV, enables us to study the vertical coherence of stagnation points. We begin with homogeneous fluid and use a phenomenological approach to prove the following theorem.

**Theorem 1**. *Isolated stagnation points of a steady pseudo-plane ideal flow in a homogeneous fluid must be vertically aligned if the corresponding pressure critical points are isolated* (*Type I, Type II* ).

If stagnation points form a continuous line and tilt vertically, the corresponding critical points of pressure, according to Lemma 1, would form an identical tilting line. Meanwhile a homogeneous fluid satisfies $p_z = 0$ and has a depth-independent pressure perturbation field. Projection of the tilting line of pressure critical points on the depth-independent pressure field results in a horizontal critical line, violating the assumption of isolated pressure critical points. Therefore the line of stagnation points can not tilt, and the above theorem is proved.

A flow without pressure perturbation, such as S1, is not subject to Theorem 1 and a tilting line of stagnation points is allowed. If the pressure critical points are not isolated, such as in S2, a skew stagnation center could occur. Theorem 1 requires concentric structure in a homogeneous flow with isolated pressure critical points, such as S6 and S9 (belong to Type I and Type II of Table III).

**Theorem 2**. *Stagnation points in a pseudo-plane flow must be vertically aligned if the corresponding pressure critical points are nondegenerate* (*Type I, Type III*).

For nondegenerate pressure critical points, $Hp \neq 0$, Lemma 3 requires the corresponding stagnation points to be isolated, and Lemma 2 further requires them to correspond to density critical points and satisfy

$$u = v = 0, \quad p_x = p_y = 0, \quad \rho_x = \rho_y = 0$$

Vertically these isolated stagnation points form a space curve

$$\mathbf{r} = x_0(z)\mathbf{i} + y_0(z)\mathbf{j} + z\mathbf{k},$$

where $z$ can be regarded as the parameter of the curve. The tangent vector to the curve is

$$\mathbf{T} = \frac{d\mathbf{r}}{dz} = x_0'(z)\mathbf{i} + y_0'(z)\mathbf{j} + \mathbf{k},$$

Since $p_x, p_y, \rho_x, \rho_y$ are all zero on this critical curve, their directional derivatives along the curve shall all be zero. For $p_x$, such vanishing directional derivative is expanded as

$$0 = \mathbf{T} \cdot \nabla p_x = x_0'(z) p_{xx} + y_0'(z) p_{xy} + p_{xz},$$

which gives

$$x_0' p_{xx} + y_0' p_{xy} = \rho_x = 0$$

Similarly for $p_y$ we obtain

$$x_0' p_{yx} + y_0' p_{yy} = \rho_y = 0$$

Because $Hp = p_{xx} p_{yy} - p_{xy}^2 \neq 0$, the above two linear equations have a unique solution $x_0' = y_0' = 0$ based on Cramer's rule, which means $(x_0, y_0)$ are independent of $z$ and the line of stagnation points is vertical. The theorem is proved.

In differential topology, a function is called a Morse function if all its critical points are nondegenerate. Theorem 2 addresses those pseudo-plane flows that have pressure perturbation in the form of Morse function, including both homogeneous and baroclinic fluids. It has overlap with Theorem 1 in Type I flows, such as S6. The baroclinic solutions S3 and S5 belong to Type III flows and their concentric structure is required by Theorem 2.

## V. Critical point analysis of quadratic flows

The above topological theory is not a complete proof of the vortex-alignment phenomenon: Type IV baroclinic flows with degenerate pressure critical points, such as S8, are not covered by either theorem. For quadratic flows, however, we are able to obtain a complete proof using an analytic approach.

Quadratic flows have the unique property of uniform strain and are commonly used in vortex analysis. The exact quadratic solutions obtained by *Sun* (2017) cover all known PIF flow patterns. Most of solution examples in Section III are quadratic. The pseudo-plane streamfunction in general quadratic form is

$$\psi = a_1(z)x^2 + a_2(z)xy + a_3(z)y^2 + b_1(z)x + b_2(z)y \tag{8}$$

Its Hessian is $H\psi = 4a_1 a_3 - a_2^2$ and a stagnation point $(x_0, y_0)$ is determined by linear equations

$$\begin{aligned} -u &= a_2 x_0 + 2a_3 y_0 + b_2 = 0 \\ v &= 2a_1 x_0 + a_2 y_0 + b_1 = 0 \end{aligned} \tag{9}$$

which can be rewritten as

$$\begin{aligned} x_0 H\psi &= a_2 b_2 - 2a_3 b_1 = A \\ y_0 H\psi &= a_2 b_1 - 2a_1 b_2 = B \end{aligned} \tag{10}$$

In the following we conduct a critical-point analysis of pseudo-plane quadratic flows and obtain three lemmas with analytical proofs.

**Lemma 4**. *A stagnation point in quadratic flows is isolated if and only if it is nondegenerate.*

Nondegeneracy is a sufficient condition for isolated critical point in the Morse lemma. For quadratic flows, we prove it is also a necessary condition. If $H\psi \neq 0$, the unique solution $(x_0, y_0)$ to Eqs.(10) is an isolated stagnation point. If $H\psi = 0$, there are two possibilities: $A = B = 0$,

Eqs.(10) have an infinite number of stagnation points that form a straightline on horizontal plane and are not isolated; $A \neq 0$ or $B \neq 0$, Eqs.(10) do not have solution and the flow has no stagnation point.

The above lemma means the isolated stagnation point of a quadratic flow can not be degenerate (see quadratic solutions S1−S7 in Table II for verification) and we only need to consider the nondegenerate type hereafter.

**Lemma 5**. *Isolated stagnation points of a steady pseudo-plane quadratic flow in the inertial reference frame must align vertically.*

Substituting streamfunction (8) into the PIF model yields the density compatibility condition (9) of *Sun* (2017)

$$2r_1 x^2 + 2r_2 y^2 + 4r_3 xy + r_4 x + r_5 y + r_6 = 0 \tag{11}$$

with coefficients

$$\begin{aligned}
r_1 &= -a_2(2a_1 a_3' + 2a_1' a_3 - a_2 a_2') + (a_1 a_2' - a_2 a_1')f \\
r_2 &= \phantom{-}a_2(2a_1 a_3' + 2a_1' a_3 - a_2 a_2') + (a_2 a_3' - a_3 a_2')f \\
r_3 &= (a_1 - a_3)(2a_1 a_3' + 2a_1' a_3 - a_2 a_2') + (a_1 a_3' - a_3 a_1')f \\
r_4 &= -2a_1(a_2 b_1 - 2a_1 b_2)' + a_2(a_2 b_2 - 2a_3 b_1)' - 2b_2(2a_1 a_3' + 2a_1' a_3 - a_2 a_2') \\
    &\quad + (2a_1 b_2' - 2a_1' b_2 + a_2' b_1 - a_2 b_1')f \\
r_5 &= -a_2(a_2 b_1 - 2a_1 b_2)' + 2a_3(a_2 b_2 - 2a_3 b_1)' + 2b_1(2a_1 a_3' + 2a_1' a_3 - a_2 a_2') \\
    &\quad + (2a_3' b_1 - 2a_3 b_1' + a_2 b_2' - a_2' b_2)f \\
r_6 &= 2b_1(a_1 b_2)' - 2b_2(a_3 b_1)' + b_2(a_2 b_2)' - b_1(a_2 b_1)' + (b_1 b_2' - b_1' b_2)f
\end{aligned}$$

For an exact solution these coefficients shall be zero everywhere. At a stagnation point $(x_0, y_0)$, Eqs.(9) give

$$b_1 = -2a_1 x_0 - a_2 y_0, \quad b_2 = -a_2 x_0 - 2a_3 y_0$$

Substituting them into $r_i = 0$ to eliminate $b_1, b_2$, we can use equality

$4 r_1 x_0 + 4 r_3 y_0 + r_4 = 0$ to obtain

$$[a_2 x'_0 - (2a_1 + f) y'_0] H\psi = 0$$

Similarly, equality $4 r_2 y_0 + 4 r_3 x_0 + r_5 = 0$ gives

$$[(2a_3 + f) x'_0 - a_2 y'_0] H\psi = 0$$

Because an isolated stagnation point in quadratic flows must be nondegenerate, i.e. $H\psi \neq 0$, the above two equations reduce to

$$\begin{aligned} a_2 x'_0 - (2a_1 + f) y'_0 = 0 \\ (2a_3 + f) x'_0 - a_2 y'_0 = 0 \end{aligned} \quad (12)$$

They have a unique solution $x'_0 = y'_0 = 0$ when the coefficient determinant is non-zero. Therefore the necessary condition for a tilting line of stagnation points is

$$(2a_1 + f)(2a_3 + f) - a_2^2 = 0 \quad (13)$$

In the inertial frame, condition (13) can not be met because the left-hand side is reduced to $H\psi$, which is required by Lemma 4 to be non-zero. This means the isolated stagnation points of a non-rotating quadratic flow must be vertically aligned, which proves the above lemma.

**Lemma 6**. *Isolated stagnation points of a non-inertial quadratic flow in the rotating frame must align vertically.*

*Sun* (2017) has proved that pseudo-plane quadratic flows in the rotating frame do not rotate with depth. As a result the *xy* term in an exact solution can be removed by coordinate rotation and we only need to consider $a_2 = 0$, in which case $r_1 = 0$ and $r_2 = 0$ of (11) are automatically met and Eqs.(12) reduce to

$$(2a_1 + f)y'_0 = 0, \quad (2a_3 + f)x'_0 = 0 \tag{14}$$

For an inertial flow with $2a_1 = -f$, $y'_0$ is allowed to be non-zero which results in a tilting line of stagnation points. For a non-inertial flow, $2a_1 \neq -f, 2a_3 \neq -f$, Eqs.(14) require $x'_0 = y'_0 = 0$ and the stagnation points must align vertically. This proves the lemma. Among the seven quadratic solutions in Section III, S1–S4 are covered by Lemma 6 and S5–S7 are covered by Lemma 5.

## VI. DISCUSSION AND CONCLUSION

This study provides both topological and analytical explanations for the concentric structure observed in the non-EB baroclinic solutions of *Sun* (2017) and the multi-polar strain-flow solutions of *Sun* (2016). A gap in the topological proof lies with the degenerate stagnation points of Type IV baroclinic flows, as Theorem 2 only deals with Morse function. Meanwhile a basic result of Morse theory says that almost all functions are Morse functions, and every function can be approximated uniformly by a Morse function (*Audin and Damian* 2014). This result is corroborated by the PIF solutions, as explained below.

Firstly, Lemma 4 prohibits streamfunction with isolated degenerate stagnation point (called non-Morse function) in a quadratic PIF solution, a fact validated by the solution set of *Sun* (2017). The depleted geometry of higher-degree polynomial solutions in *Sun* (2016), which contains three forms, further suggests that circular vortex like S8 is the only candidate for a non-Morse baroclinic solution. (Among the other two forms, straightline jet does not have isolated stagnation point and multipolar strain field is barotropic.) Because baroclinic circular vortices are always EB according to Theorem 2 of *Sun* (2008), we conclude that the gap in the topological proof is dynamically amendable. In other words, vertical alignment is expected for all four types of isolated stagnation points in Table III, except for those in inertial flows.

Because Theorem 1 involves no PIF dynamics and Theorem 2 only uses the hydrostatic equation, their application is well beyond steady ideal flows. For example, Theorem 1 can be applied to a viscous flow as long as its stagnation points correspond to pressure critical points. In Appendix A, the circular-vortex case of the viscous von Kármán swirling flow is analytically solved, showing that the vertical line of vortex center is curvilinear rather than a straightline. The result can be explained by the topological theory: if the line is straight, the horizontal pressure gradients at the vortex center would be zero, in which case Theorem 1 is applicable and the tilting vortex is prohibited.

The fact that vertical alignment exists in homogeneous, stratified, rotating and non-rotating fluids alike, suggests that the phenomenon is not due to stratification or rotation effect. Instead, streamwise conservation of vorticity and density in a pseudo-plane ideal flow, coupled with the hydrostatic balance, leads to the alignment of vortex center with the direction of gravity. It reminds us of the assertion by *Arnold and Khesin* (1998, page 50) that the ideal fluid dynamics is analogous to the classical theory of rigid body with a fixed point. The topological theory of vertical alignment is established here for steady pseudo-plane flows. Evolution to such a steady state, however, involves transient instabilities and shall be examined in future study.

**Acknowledgement**

This work was supported by the National Natural Science Foundation of China (Grant No. 41576017, 41421005). The encouragement and comments from three anonymous reviewers are greatly appreciated.

**Appendix A. Tilting circular vortex in viscous fluid**

For a viscous incompressible fluid between two parallel disks rotating around different axes at the same angular velocity, *Berker* (1982) obtained a steady pseudo-plane solution without vertical circulation, which can be regarded as an equilibrium state of the well-known von Kármán swirling flow (*von Kármán* 1921, *Batchelor* 1951). In engineering application, the Berker solution is related to the flow in orthogonal rheometer (*Pucci and Saccomandi* 2017) and a simple derivation of this viscous exact solution is provided here.

For steady pseudo-plane flows in a viscous incompressible fluid, the Navier-Stokes equations become

$$uu_x + vu_y = -p_x + \nu \nabla^2 u$$
$$uv_x + vv_y = -p_y + \nu \nabla^2 v \tag{A1}$$

where $\nu$ is kinematic viscosity. The pressure perturbation in a homogeneous fluid has to be vertically invariant, i.e.,

$$p_z = 0 \tag{A2}$$

Assuming a tilting circular vortex in quadratic form

$$\psi = \frac{\omega}{2}[(x - x_0(z))^2 + (y - y_0(z))^2]$$
$$u = -\omega[y - y_0(z)], \quad v = \omega[x - x_0(z)] \tag{A3}$$

and substituting it into (A1), we obtain pressure perturbation

$$p = \frac{1}{2}\omega^2(x^2 + y^2) + (\nu y_0'' - \omega x_0)\omega x - (\nu x_0'' + \omega y_0)\omega y \tag{A4}$$

For (A2) to hold, the coefficients of the linear terms in (A4) have to be vertically constant, i.e.,

$$\nu y_0'' - \omega x_0 = a, \quad \nu x_0'' + \omega y_0 = b \tag{A5}$$

where $a$ and $b$ are constant. From (A4), the pressure field is circular with vertically invariant center at $(-a\omega^{-1}, b\omega^{-1})$. This pressure center clearly differs from the stagnation center $(x_0, y_0)$.

For an ideal fluid, $\nu = 0$, (A5) requires $x_0$ and $y_0$ to be constant, which means the circular vortex is vertically concentric. For a viscous fluid, $\nu \neq 0$, (A5) produces two fourth-order ODEs

$$\nu^2 \frac{d^4 x_0}{dz^4} + \omega^2 x_0 = -a\omega, \quad \nu^2 \frac{d^4 y_0}{dz^4} + \omega^2 y_0 = b\omega$$

which have a general solution

$$x_0 = c_1 e^{mz} \cos(mz) + c_2 e^{mz} \sin(mz) + c_3 e^{-mz} \cos(mz) + c_4 e^{-mz} \sin(mz) - a\omega^{-1}$$
$$y_0 = c_1 e^{mz} \sin(mz) - c_2 e^{mz} \cos(mz) - c_3 e^{-mz} \sin(mz) + c_4 e^{-mz} \cos(mz) + b\omega^{-1}$$

(A6)

Here $c_1, c_2, c_3, c_4$ are arbitrary constants and $m = \sqrt{\omega/2\nu}$. A horizontal translation of origin would eliminate the constant $\omega^{-1}$ terms in (A6) and move the pressure center to the origin. In an orthogonal rheometer, two parallel plates rotate around distinct axes at $(x_1, y_1)$ and $(x_2, y_2)$ with the same angular velocity $\omega$. Thus we have four linear algebraic equations to determine the values of four constants in (A6). The resultant spatial line of circle center $(x_0, y_0)$ is curvilinear rather than a straightline.

The solution demonstrates that circular vortex with skew center is permitted in viscous fluid. The non-straight tilting line is also explained: at the stagnation center $(x_0, y_0)$, pressure gradients derived from (A4) are

$$p_x = \nu\omega y_0'', \quad p_y = -\nu\omega x_0''$$

and yield $p_x = p_y = 0$ if the spatial line $(x_0, y_0)$ is straight. In that case Theorem 1 becomes applicable and requires the spatial line to be vertical, which is contradictory.